# THE SPACE OF ALL PATHS FOR A QUANTUM SYSTEM: REVISITING EPR AND BELL'S THEOREM

Warren Leffler
Department of Mathematics,
Los Medanos College,
2700 East Leland Road Pittsburg,
CA 94565
wleffler@losmedanos.edu

ABSTRACT

In this paper we identify a hidden premise in Bell's theorem: measurability of the underlying space. But our system (the space of all paths, SP) is not measurable, although it replicates the predictions of standard quantum mechanics. Using it we present three counterexamples to Bell's theorem and also show why Bell-like arguments for more than two particles cannot be carried out in this model. Moreover, we show that the result places severe constraints on possible viable interpretations of quantum mechanics: Either an interpretation must in some form represent a quantum system in terms of all paths within the system or, alternatively, the interpretation must harbor "action at a distance."

## I. INTRODUCTION

**A. A tacitly assumed premise in Bell's theorem::** The philosopher Tim Maudlin has a very useful formula for characterizing various failed attempts to refute Bell's theorem: "No local *X* theory can make The Predictions [the quantum mechanical predictions] for the results of experiments carried out very far apart." Assuming Bell's theorem is correct (as Maudlin argues), the X is superfluous. He calls this "The Fallacy of the Unnecessary Adjective." Maudlin also states: "… if Bell unknowingly employed a hidden premise in the derivation of his famous inequality and consequently misunderstood what he had proven [it would] … force a radical reevaluation of Bell's work. However, Bell employed no such hidden premise …."[1]

The interesting thing, though, is that all proofs of Bell's theorem (his original arguments and those by others in the same vein) [2] for two entangled particles involve a probability distribution. This means that there is indeed a hidden premise, a tacitly assumed "X"—namely, that the underlying space for a quantum system is measurable. In other words, if we choose "X" to be "measurable" then in Maudlin's formula we have the proposition, "No local, measurable theory can make The Predictions for the results of experiments carried out very far apart." We consider Bell's simple proof of this specific proposition (that is, when "measurable" is substituted for X) to be obviously valid.

**B. Space of all paths:** In this paper, however, we develop a new and (so we believe) indisputably elegant approach to quantum mechanics, "the space of all paths" (SP). Although it is new, SP is in effect a synthesis of Richard Feynman's formulation with a model of particle interaction described by David Deutsch, [3] a model that Deutsch based on Hugh Everett's many worlds interpretation, MWI. SP replicates the predictions of standard quantum mechanics but, unlike MWI, it is a system in which there is a single outcome for each quantum event. This fact makes it an appropriate space (again unlike MWI) for examining both Bell's theorem for two entangled particles and also extensions of the theorem to more than two particles.

**C. Space of all paths is not measurable:** As is well known, however, the space of all continuous paths joining two points in coordinate space is not measurable.[4] But here is a quick sketch showing there can be no countably additive and translation invariant measure on an infinite-dimensional space such as SP. Suppose there were, say, such a Lebesgue measure $\mu$, where $e_k = (0,...,0, \underset{k\text{th place}}{1}, 0,...)$ is an orthonormal basis for the space and $B$ is a sphere of radius 2 centered at the origin. For each $k$, form the sphere $B_k$ of radius ½ centered at $e_k$. Then the spheres $B_k$ form a disjoint class, and so we see that we have a contradiction (because a Lebesgue measure $\mu$ must be finite on bounded Borel sets, additive, and translation invariant):

$$\mu(B) \geq \sum \mu(B_k) = \frac{1}{2} + \frac{1}{2} + \frac{1}{2} + \ldots = \infty.$$

A probability distribution (and therefore a countably additive measure) is in fact essential to all proofs of Bell's theorem for two entangled particles. Still, it is perhaps not surprising that no one until now has noticed the suppressed premise of measurability in Bell's argument, since in learning quantum mechanics virtually everyone is trained initially in some form of von Neumann's Hilbert-space approach to the foundations. Indeed, here is a statement from David Griffiths's excellent introductory text: "… such non-normalizable solutions cannot represent particles, and must be rejected. Physically realizable states correspond to the square-integrable solutions to Schrödinger's equation."[5] Measurability is of course presupposed in such approaches to the foundations.

**D. Failure of Bell-like extensions to more than two particles:** There are also Bell-like theorems that involve more than two particles, for example that of Green, Horne, and Zeilinger, (GHZ).[6] Interestingly, however, such arguments are relatively easy to overturn in SP (Sec. IV).

**E. EPR criteria:** The GHZ extension (in the above cited paper) directly refers to a famous paper co-authored by Einstein, the EPR paper of 1935, which argued that quantum mechanics fails to provide a complete description of physical reality.[7] In this paper we will also refer to the following four conditions or criteria that EPR claimed must be met by any complete theory. In describing these conditions, EPR focused on position and momentum, but these have later been recast conveniently in terms of two entangled spin ½ particles as follows:[8]

(i) *Perfect correlation*: If the spins [of the two particles] are measured along the same direction, then with certainty the outcomes will be found to be opposite.

(ii) *Locality*: Since at the time of measurement the two systems no longer interact, no real change can take place in the second system in consequence of anything that may be done to the first system.

(iii) *Reality*: If, without in any way disturbing a system, we can predict with certainty (i.e., with probability equal to unity) the value of a physical quantity, then there exists an element of physical reality corresponding to this physical quantity.

(iv) *Completeness*: Every element of the physical reality must have a counterpart in the [complete] physical theory.

**F. Distinction between a mathematical representation and the phenomena represented:** For the discussion in this paper we will now emphasize an important distinction between what might be called "quantum events" (the underlying experimentally observed phenomena) and a mathematical theory for predicting the observed events. By standard quantum mechanics (SQM) we mean a mathematical representation of quantum phenomena generally based on some form or other of Lebesgue square-integrable wavefunctions over a separable, Cauchy-complete Hilbert space. The space of SQM functions is often supplemented by so-called delta functions, and there are versions of SQM based on a rigged Hilbert space, and so forth. But these differences among versions of SQM are not important for our discussion.

What is important for our discussion is that SP is an accurate mathematical representation of quantum phenomena, equal to that of SQM, [9] although the underlying space (space of all paths) is not itself measurable. That is, as is well known, SP replicates the predictions for quantum observables (physical quantities whose values can be measured by an experiment).

**G. The SP mathematical representation:** Given two points in a quantum space, SP quantifies the amplitude for a single tangible particle (see Sec. II) to travel between the points by the sum or integral of "exponentiated-action" terms, $\exp(iS[x(t)]/\hbar)$, over all possible paths $x(t)$ between the points—where $S[x(t)]$ is the action over a path $x(t)$. For position/momentum, the paths are in $\mathbb{R}^3$ or in a topologically constrained subspace of $\mathbb{R}^3$, where the action over a path for position/momentum is the integral of the Lagrangian (difference of kinetic and potential energy) at each point on the path. [10] For spin the paths are "rotational" paths in $\mathbb{R}^3 \times SO(3)$, where $SO(3)$ is the group of rotations of $\mathbb{R}^3$ with determinant 1. In the case of spin, the action is a sum at each instant along a rotational path associated with a "top" action and a magnetic-field action. [11] The amplitude for a particle traveling between two points $x_1$ and $x_2$ in a space (whether for coordinate or spin space) is proportional to the time-development operator or propagator

$$K(x_2,t_2;x_1,t_1) = \sum_{\substack{\text{over all paths} \\ \text{from } x_1 \text{ to } x_2}} \exp(iS[x(t)]/\hbar) \tag{1}$$

In this paper, however, we will never actually need to calculate a path sum or integral, in effect taking as an axiom the well-known fact that the path-integral approach, both in coordinate and spin space, predicts the experimental outcomes.

In the usual accounts of the path-integral formulation it is frequently stated that the particle (tangible particle, see below) in traveling between two points pursues all possible paths (as Feynman put it, "[A particle] smells out all paths in the neighborhood and chooses the one that has the least action…").[12,13] This would of course be a paradox, since the particle would have to travel at superluminal speeds, even when pursuing all paths within a small, finite region of space around the classical path (since there uncountably many paths in such a region). But this is easily resolved in Sec. II.

**II. HIDDEN REALITY**

**A. Random path postulate and the shadow stream**: In SP all possible paths in a quantum space are traveled in a one-to-one fashion by actual entities or particles—that is, traveled by "shadow" and "tangible" (ordinary) particles. The concept of a "shadow particle" comes from Deutsch's qualitative analysis of single-particle interference in the two-slit experiment (which, as noted above, Deutsch based on MWI). In his discussion Deutsch argued that when an ordinary particle (a tangible particle) travels between two points the interference effects are the result of interactions with counterpart shadow particles. [14] These counterpart particles interfere only with tangible particles of the same type, and therefore they can be detected only indirectly, through their effects on regular, ordinary particles—that is, shadow photons interfere only with regular photons, shadow electrons only with regular electrons, and so on.

Thus, following up on Deutsch's concept, we postulate in SP that whenever a single tangible (ordinary) particle travels from one point to another in a space (whether for position/momentum, $\mathbb{R}^3$, or for spin, $\mathbb{R}^3 \times SO(3)$—see below) and has a choice of paths that it can take, it randomly takes one of the possible paths, and distinct shadow particles take the others. This is called the "random path postulate," RPP. In other words each tangible particle generates a stream of accompanying counterpart particles, called the tangible's "shadow stream" (we include the tangible itself in the shadow stream).

Drawing on the hyperreal number system (Abraham Robinson's rigorous approach to infinitesimal and infinite quantities—nonstandard analysis),[15] we postulate further that, although the paths in a shadow stream generated by a tangible particle can be infinitely close to each other and interfere as waves interfere, they never intersect. As a helpful heuristic picture, think of the particles traveling along the various intertwined paths as being in different but related worlds *à la* Everett. Also we will suppose that shadow particles obey the same dynamics (for example, the Lagrangian) as tangible particles.

**B. NIP (non-interacting tangible particles):** For two non-interacting *tangible* particles the composite amplitude of the two streams generated by the tangibles is the product of the separate amplitudes. (Here we are simply using Feynman's rule that for two non-interacting tangible particles "the amplitude that one particle will do one thing and the other one do something else is the product of the two amplitudes that the two particles would do the two things separately."[16])

**C. A non-measurable space but a scientifically valid theory:** Although the underlying space in SP (space of all paths) is not measurable, the SP theory on which it is based is nevertheless a valid theory, since (as is well known) the sum of exponentiated-action terms over paths in the space converges to the standard amplitude (for example, the amplitude obtained from Schrödinger's equation). There is, as yet (so far as we know), no general proof that the path-formulation always coincides with the SQM result, although there are by now a great many examples that have been accomplished by decades of deep mathematical research since Feynman's original result.[17] Moreover, in practice one is usually concerned with just a satisfactory approximation, and to that end chooses a cut-off region for the integral. In some respects the situation is analogous to the theory of the Riemann or the Lebesgue integral, in which many integrals for a wide variety of integrands can be given in a closed form, although there are equally many integrands that have no closed form (although the conjecture for the path integral is that it always give the same result as in SQM).

Also, one should keep in mind that the alternative, a measurable space, leads to Bell's result that widely-separated parts of the universe are tied together by instantaneous influences according to a dynamic that no one understands.

**D. A sum of unit vectors:** As we see in the figure below the convergence of the path integral is, heuristically, a result of wild oscillations of unit vectors in the complex plane, the oscillations tending to "cancel" each other (pointing in opposite directions in the sum) except in the neighborhood of the classical path where the vectors add up constructively—not at all the sort of thing envisaged in Bell's argument involving entangled particles.

Fig. 1 illustrates how this cancellation occurs in a simple example: when a single tangible particle (accompanied by shadow particles) travels from a source, $S$, to a mirror and then is reflected up to a reception point, $C$.[18] Later, in Sec. V, we will see that this same configuration occurs as a part of one of our counterexamples to Bell's theorem, when the counterexample is based on a standard two-particle Rarity-Tapster interferometer.

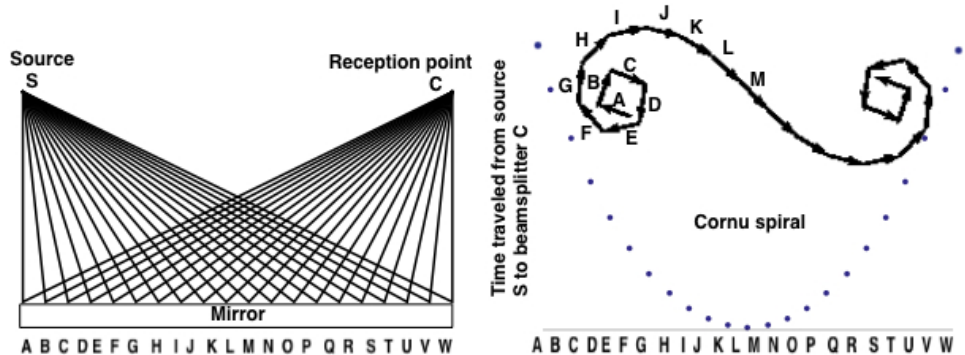

Figure 1

A free tangible particle is emitted from a source, S, randomly taking one of the paths from S to a mirror and then up to a reception point, C. Shadow particles in SP travel the other paths, since the tangible generates an infinite stream of shadow particles going to the mirror and them up to C. The space of all paths is not measurable, but (as is well known) the sum of exponentiated action terms (the unit vectors A, B, C, ..., which are depicted in the Cornu spiral) yields the amplitude (whose absolute square is the probability that the tangible will arrive at C).

**E. Infinitely-many-to-one mapping from SP onto SQM, and the Relation of SP to SQM:** In a letter to Born, Einstein wrote:[19]

> Assuming the success of efforts to accomplish a complete physical description, the statistical quantum theory would, within the framework of future physics, take an approximately analogous position to the statistical mechanics within the framework of classical mechanics. I am rather firmly convinced that the development of theoretical physics will be of this type; but the path will be lengthy and difficult.

The mathematical relationship of SP to SQM is precisely of the kind just described by Einstein—the interaction of shadow and tangible particles generating the observables of SQM, the observables arising in a probabilistic fashion.

There is in fact an infinitely-many-to-one homeomorphic mapping of SP onto SQM in which homotopy classes of paths are mapped onto orthogonal basis vectors in the Hilbert-space representation.[20] (Recall that two paths with the same end points are homotopic when they can be continuously deformed to each other). The mapping is not unique, and in Sec. III we present examples of several obvious mappings in our further discussion of SP (Fig. 1 above illustrates a particularly trivial such mapping).

**F. Opaqueness of SQM to the mechanism underlying quantum observables**: Although SQM is a highly accurate and elegant computational framework for computing quantum amplitudes, it is utterly opaque—unlike SP (as we show below)—to the underlying mechanism that produces the observables. (Moreover, it is littered with single-particle interference paradoxes that no longer exist in SP, although discussing these is beyond the scope of the present article.) It is true that the calculations in SQM are generally easier to do than in SP, but this is partly because it is generally easier to solve a differential equation (say, Schrödinger's equation) than to find the closed form of an integral; but there is a further reason, stemming from the fact of the infinitely-many-to one mapping of II.E above. Indeed, it is usually easier to calculate in an infinitely-many-to-one homomorphic image than in the pre-image. As a crude but perhaps insightful analogy, compare the integers to the integers mod *n* (a homomorphic image of the integers); in the latter space, say mod 5, it would be is easy for a child to learn the rules of calculation while being completely unaware of the larger pre-image of the integers.

## III. IN SP ELEMENTS OF REALITY ARE NOT THE SAME AS EIGENVALUES OF SELF-ADJOINT OPERATORS

**A. Element of reality:** To be sure, Einstein *et al.* did not precisely define what they meant by "element of reality" (condition (iii) in Sec I.E above), though one could quite reasonably interpret their discussion of a particle's position and momentum in classical terms. In any case, in SP an "element of reality" is a classical property, such as the classical momentum or classical spin for a tangible particle in the stream. Also, particles are never in superposition as is typically the case in, say, the Copenhagen interpretation, although the shadow stream itself might be (since it is analogous to a wavelike stream).

Taking this view makes perfect sense in SP but not in SQM, because in SP quantum events are a sum of interference effects involving a tangible particle with (usually uncountably many) classically moving counterpart particles—the particles traveling in coordinate space or in spin space ($\mathbb{R}^3 \times SO(3)$). Each particle in the collective entity of the shadow stream obeys classical dynamics and has a corresponding classical property as it travels along its respective path in the stream.

**B.1 Particle on a ring:** As an example of an infinitely-many-to-one mapping from SP onto the SQM Hilbert space, consider a frequently treated problem in elementary quantum mechanics (which we need also in Sec. V)—that of determining the momentum of a so-called quantum particle-on-a-ring (roughly pictured by an electron moving on a conducting ring in, say, benzene). Using Schrödinger's equation (and taking momentum to be the Fourier transform of position, etc.) one shows in SQM that there is a discrete set of observables—a denumerably infinite set of eigenvalues for momentum (in contrast to what occurs with a classical bead traveling around a frictionless ring where, in principle, we can obtain any of a continuum of momentum outputs).

**B.1.a Path-integral approach:** The particle-on-a-ring problem is also solved easily using the path integral.[21] Thus assume that for the ring we have a free tangible particle constrained to move on a circle, the particle able to go any number of times back and forth along the circle (as noted in Sec. II.A above, all paths in SP are infinitesimally separated from each other and do not intersect spatially, although they may interfere at each instant when corresponding values are added). Although each particle in the tangible's shadow stream travels one of a continuum of paths around the ring, each path is associated with a winding number, an integer *n*—that is, the number of times the particle travels past a given fixed point in the positive direction (counterclockwise) minus the number of times it passes the point in the minus direction (clockwise).

**B.1.b Propagator as a linear combination of propagators over homotopy classes:** We can input a possible continuum of momentum values to the ring particle, but when we measure the eventual output we always obtain only one of a pre-determined, denumerably discrete set of eigenvalues in a one-to-one correspondence with winding numbers. Moreover, unlike for Euclidean space—where Feynman's formulation was originally developed—the associated configuration space for the ring is not simply connected, a consequence of the topological constraint of confining the tangible particle's path to a circle. This means that the path-propagator for the ring is a linear combination of propagators over each homotopy class in the space,[22] the classes corresponding to winding numbers

(which illustrates an obvious mapping from SP onto SQM as mentioned in Sec. II.D).

**B.1.c Uncountably many path configurations in SP associated with a discrete set of eigenvalues for the momentum operator.** As always in SP, the underlying space of all possible paths (paths around the circle) is *not measurable*. Nevertheless, for the reason illustrated in Fig. 1, the sum of uncountably many exponentiated-action terms over the many possible associated paths for the winding number always yields a momentum value lying in the discrete set of eigenvalues predicted by Schrödinger's equation in the Hilbert-space formulation of SQM. In SP the tangible particle's measured value (the eigenvalue of an operator in SQM) is an outcome of interference effects with the accompanying shadow particles in its homotopy class. Therefore within a winding-number (homotopy) class there are infinitely many paths that the tangible particle could take (each path having a corresponding exponentiated-action term), although by RPP in Sec. II.A the tangible randomly takes just one of the paths. Any two actions (*S*[*path*]) with the same winding number can in principle be switched in the sum and yet lead to the same value when summed with exponentiated-action terms from other paths. In other words, at the time of measurement the tangible's element of reality *is associated with exactly one of these paths*, but the sum of exponentiated-action terms over all such paths is the observable (the eigenvalue of the standard operator). Thus the tangible particle has a certain momentum at the time of measurement (*clearly what EPR would call the associated element of reality in this context*), but there is only a vanishingly small chance that a measurement will actually exhibit that particular value.

Thus in SP infinitely many configurations of paths (associated with distinct elements of reality) are mapped to each particular eigenvalue in SQM for a given operator. This is what underlies the characteristic discreteness of quantum observables in general, not just those regarding a quantum bead. Indeed (referring to EPR's conditions in Sec. I.E), it is why SQM is not a complete theory of physical reality.

**C. Spin 1/2** Here, as a further example, is how the mapping from SP to SQM works for spin. In SQM the experimental results for spin are elegantly described using a finite dimensional vector space over the complex numbers. In fact for spin-½ the mathematical description seems simplicity itself, perhaps the simplest nontrivial Hilbert space of all: a two-dimensional vector space over the complex numbers. Nevertheless, although the vector algebra predicts the observed outcomes with amazing precision, it is totally opaque as to what is actually going on behind the observed phenomena—phenomena that resemble classical spin but differ in various well-known ways.

**C1. A classically spinning tangible particle in a stream of accompanying shadow counterparts:** In SP we postulate that a spinning tangible particle is a tiny spinning top that travels in $\mathbb{R}^3 \times SO(3)$ within a shadow stream of other such tops (which we can think of also as tiny dipole magnets), each particle in the stream obeying the classical dynamics of spin.[23] We can then account easily for the outcomes of the small number of fundamental Stern-Gerlach (SG) experiments that form the basis of the standard Hilbert-space representation of quantum spin.[24] This account is an application of a key property of *SO*(3): namely, that it has exactly two homotopy classes of paths, and hence that each particle in the stream is traveling in one of the two classes. Thus, suppose a tangible electron and its shadow counterparts enter the inhomogeneous field of a SG device. We postulate that if the tangible exits a prior SG field and then enters a field of the same orientation it will continue in that homotopy class. But if it enters a field oriented in a different direction it will go randomly (by RPP, the random path postulate) into one of two homotopy classes for the new orientation.

Indeed, the tangible particle and some shadow particles in the tangible's homotopy class may initially have larger spin components in the direction set by the SG device. (This would also enter into an SP homotopy account of other kinds of spin particles, such as spin 1, where (in addition to those of *SO*(3)) there are more than two homotopy classes—but we confine our discussion in this paper to spin 1/2). Thus, depending on which homotopy class it is in, we picture each particle in the class as a spinning top disposed by the SG field to land more "up" or more "down" on a detecting plate, but where the tangible is observed to land depends on the interference effects coming from the accompanying shadow tops in the tangible's homotopy class. The tangible's final position on a detecting plate when it leaves the SG field is an outcome of the sum of interference effects within its homotopy class, producing a discrete set of observed positions (which, as noted above, is generally the case in quantum mechanics).

In particular, a spin-½ tangible particle (as in Fig. 2 below) will be detected in either of two positions, depending on the homotopy class it is sent to by the SG magnetic field. Each particle in the tangible's homotopy class has a classical spin, quantified by the exponentiated action over the particle's "spin path" in $\mathbb{R}^3 \times SO(3)$. At the time of measurement any two of the infinitely many paths in the tangibles homotopy class could be switched in principle, the tangible randomly traveling on either path, and yet produce the same total sum. In the case of spin ½, for example, each such path configuration will lead either to the position of $+\hbar/2$ or $-\hbar/2$ on the SG detection plate. For example in the finite snapshot of Fig. 2 we could interchange the "darker" dipole magnet (signifying a tangible particle) with any of the other paler particles in the tangibles homotopy class. In this way, uncountably many spin configurations (each a potential element of reality) are thus mapped to a discrete set of outcomes.

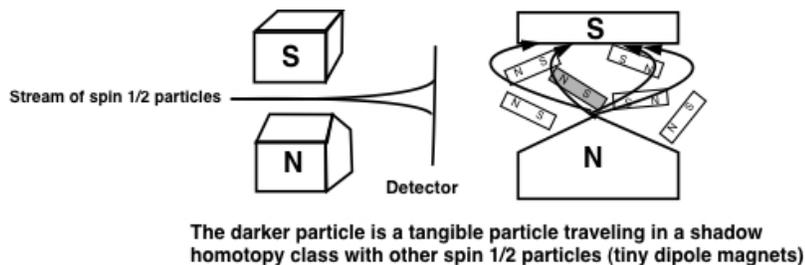

The darker particle is a tangible particle traveling in a shadow homotopy class with other spin 1/2 particles (tiny dipole magnets)
Fig. 2

But what we observe when we perform a measurement is what is called in SQM the particle's "intrinsic spin," which differs in many ways from classical spin. For example in SQM non-commuting operators have no common eigenvalue. Thus in SQM a particle does not, say, possess simultaneous *x* and *y* intrinsic spin components. What we observe of intrinsic spin is always one of a fixed discrete set of eigenvalues for each type of tangible particle. But in SP these observed values are just the sum of interference effects involving the tangible's interaction with the other particles traveling in the tangible's homotopy class. At each instant the tangible could be spinning randomly in any of infinitely many possible directions, with the total sum of interference effects among all particles in the shadow homotopy class producing the SQM observable. Thus at each instant the element of reality for a tangible particle is that of a classical spinning top in a stream of classically spinning counterpart shadow particles. The measured result at that instant is the sum of exponentiated-action terms over all possible spin-paths in the homotopy class. In other words, infinitely many possible elements of reality are always mapped to one of the finitely many eigenvalues. To reiterate: In SP eigenvalues are not the same as elements of reality. This distinction is paramount in our answer in Sec. IV to GHZ's challenge to the principles of EPR.

**D. Each element of reality having a counterpart in the theory:** In SP each element of reality is associated with a tangible particle traveling a particular path (as in condition (iv) of Sec. I.E). On the other hand, in SQM only the *extractable* information is available. What is left out is typically infinite, with every one of the usually infinite number of paths in SP corresponding to a distinct physical state of the system. Indeed, we know that each distinct *preparation* procedure in SQM corresponds to an input of information to a quantum system, a distinct wavefunction—usually an uncountable infinity of possibilities. However, the output, the *extractable* information in SQM—because of its relation to orthogonal wavefunctions in the infinitely-many-to-one mapping from SP to SQM—corresponds to only a small fraction (indeed, only a vanishingly small fraction) of the system's *total* information in wavefunction form.

**IV. THE FAILURE OF GHZ IN SP**

We shall now show that GHZ fails in the space of all paths. To this end we analyze David Mermin's splendidly clear version of GHZ's three-particle argument.[25]

In his inimitable style Mermin asks (facetiously), "What's wrong with these elements of reality?"— challenging the reader to find something wrong with them. Well that's easy to do, in light of the discussion in Sec. III. Indeed, GHZ assumes that EPR's "elements of reality" are eigenvalues of self-adjoint operators. Thus in his account of GHZ, Mermin derives six of what he calls "elements of reality," $m_x^i, m_y^i$, where the three particles are named 1, 2, and 3, and the subscripts *x* and *y* refer to corresponding Stern-Gerlach measurements. Now EPR's conception of reality is based on principle (iii) in Sec. I.E: "If, without in any way disturbing a system, we can predict with certainty the value of a physical quantity, then there exists an element of physical reality corresponding to this physical quantity." Thus Mermin states: "All six of the elements of reality have to be there, because we can predict in advance what any one of the six values will be by measurements made so far away that they cannot disturb the particle that subsequently does indeed display the predicted value." Yes, that is certainly true in SP as we show in Sec. V, although in his discussion Mermin is only assuming it for the sake of argument; but he next assumes that these elements of reality are eigenvalues, the numbers +1 or –1 (for spin up or down in units of $\hbar/2$). He then forms a certain product of them and concludes that "the product of the three resulting values must once again be +1." But, as he notes, this is immediately seen to be contradicted in a particular instance by the quantum mechanical prediction. Of course his argument fails because, to put it simply, although one can multiply numbers, one cannot in effect meaningfully multiply path configurations. That is, the GHZ argument cannot be carried out in SP because, as we saw above, eigenvalues of quantum operators are not the same as elements of reality: they are simply the image of a sum of interference effects over all possible paths, with the tangible particle randomly taking one of the possible paths. QED

The GHZ argument has been called "ingenious" and "strikingly powerful," because it is not statistical in nature, like Bell's theorem, but supposedly refutes EPR by the outcome of *one* experiment consisting of a single run. Yet, in a way, the proof merely amounts to showing, "If EPR is wrong (in other words, if SQM is right), then EPR is wrong."

**V. COUNTEREXAMPLES TO BELL'S THEOREM IN SP**

In explaining his theorem Bell once assumed, for the sake of argument, that there was a strong correlation in frequency of heart attacks in the two French cities of Lille and Lyons.[26] He then pointed out that the probability of *M* cases in Lille and *N* in Lyons are not independent—that is, they do not separate: $P(M, N) \neq P(M)P(N)$. But, as he further explained, if we let $\lambda$ represent all the so-called "hidden variables" or "common causative factors" and use conditional probability, then we have a residual equation that expresses independence and therefore local causality: $P(M, N \mid \lambda) =$ $P(M|\lambda)P(N|\lambda)$. He continues: "Let us suppose then that the correlations between [the measurements on the two sides in the EPR experiment] are likewise 'locally explicable'. That is to say we suppose that there are variables $\lambda$, which, if only we knew them, would allow decoupling of the fluctuations": $P(\alpha,\beta|\lambda) = P(\alpha|\lambda)P(\beta|\lambda)$ (here we have slightly—though immaterially—modified his original notation and used $\alpha$ and $\beta$ to represent the measurements).

From this assumption of local causality in the form of independence or "decoupling of the fluctuations," he then derived a simple inequality that is violated by the predictions of quantum mechanics. He thus concluded that the correlations for a pair of entangled quantum particles emitted from a source are coordinated by "action at a distance," an influence that communicates instantaneously across the origin. This hidden influence or communication is now widely accepted as fact. For instance in a report regarding a famous recent interferometer experiment in Geneva (famous because the measurements were separated by more than 10 km), the researchers ask: "How can these spatially separated locations 'know' what happens elsewhere? Is it all predetermined at the source? Do they somehow communicate? … *that it is all predetermined at the source, has been ruled out by numerous Bell tests* [italics added] … If some sort of hidden communication does exist, it must propagate faster than light … the hypothetical hidden communication must propagate at least $10^7$ times faster than the speed of light!"[27, 28]

Indeed, if the underlying space is measurable then Bell's simple mathematical derivation (not any experimental "Bell test") establishes a result of incredible implausibility. As mentioned earlier, the superluminal "communication" above would have to take place according to mysterious laws of dynamics that no one understands; it is in fact so implausible that one could almost—despite the history of the theorem's widespread acceptance—take it as a *reductio ad absurdum* argument in favor of non-measurability of the space.

In any case, the space of all paths goes "outside the box" of the standard setup and provides (in terms of congruent paths common to the two sides of the experiment) information about the paths not traveled by the tangibles when the shadow particles contiguously interact with the tangible particles. The interaction occurs for spinless particles in a two-particle interferometer at a beamsplitter; for spin it takes place in a Stern-Gerlach magnetic field, where the spin paths are in SO(3) (the rotations of three-space with determinant 1). There are indeed "variables $\lambda$" that allow "decoupling of the fluctuations." As we shall show, they are the exponentiated-action terms corresponding to the congruent paths common to the two sides of the experimental setup.

**A. Counterexample 1: Spinless particles in an interferometer**: In our first counterexample to Bell's theorem we will discuss a standard experimental setup for demonstrating two-particle correlations, showing that Bell's theorem fails in this instance—thereby, overturning the general theorem.

Except for the labeling, our interferometer (a Rarity-Tapster interferometer) is identical to the one analyzed in an important paper by Bernstein, Green, Horne, and Zeilinger (BGHZ): "Bell theorem without inequalities for two spinless particles."[29] For their proof to go through, BGHZ found it necessary to augment the proof with a premise called "emptiness of paths not taken," EPNT. Interestingly, in discussing their premise, BGHZ actually (no pun intended) foreshadowed our approach:

> … one can deny EPNT and thereby imagine that *something* could travel down the empty beam, so as to provide information to the nonempty beam, when the two beams meet. And this something could be consistent with EPR locality, if the particles (and these somethings) on opposite sides of the origin do not communicate.

Of course we deny EPNT in the space of all paths, a fact that we will return to later in Sec. VI to establish a result that has profound consequences for quantum foundations. (We do this by extending the BGHZ argument to homotopy classes of paths in a general, multiply-connected space in place of the small number of paths discussed in the BGHZ paper). But for now notice that the interferometer diagram on the left in Fig. 3 below indicates only four paths from the source to the beamsplitters—which is the conventional picture from BGHZ. In SP, however, this is not the actual space of paths involved, which is pictured on the right (in a partial form of course), in the fashion of Fig. 1.

In the right-most drawing, moreover, the paths are meant to be in three dimensions. For two-particle correlations to occur in the kind of interferometer that we are considering, it is a well-known experimental fact that the source, $S$, cannot be a point source, as in single-particle interferometer experiments.[30] *The source must have positional uncertainty—three degrees of freedom*. Because of this the possible trajectories of particles issuing from the source are handled by ceiling and floor mirrors (M-upper, M-lower, etc.). The particles traveling, for example, along the upper paths are reflected down from *ceiling* mirrors and then on to the beamsplitters (at $C$ and $C'$). From the beamsplitters they finally go on to detectors ($u$, $d$, etc.). In the left-most, two-dimensional picture the upper paths are labeled $a$ and $b$ on the left-side and $a'$ and $b'$ on the right-side of the interferometer. On the other hand, in the right-most, three-dimensional diagram the paths belong to homotopy classes $A, B, A', B'$ containing infinitely many paths. There are phase shifters $\alpha$ and $\beta$ along the upper paths. The right-most figure displays some of the paths in the actual space of all possible paths in a region of $\mathbb{R}^3$. This is an infinite-dimensional, functional space, and is therefore *not measurable*.[31]

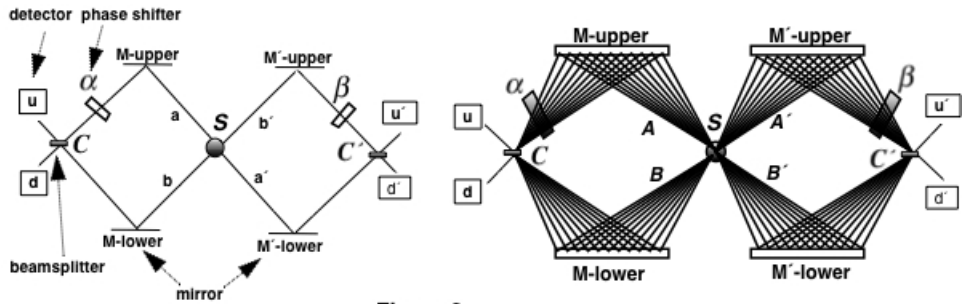

Figure 3

Of course if the quantum space were measurable—as everyone has hitherto assumed—then Bell's theorem (the result of a perfectly valid mathematical proof based on the premise of measurability) would apply and there would be no counterexample. But it is easy to show that we do in fact have a counterexample, using an idea hinted at (unintentionally) in BGHZ's above comment about the denial of EPNT.

We also note that it is an experimental fact for our type of interferometer that the emission angle subtended at the source by the upper paths must be sufficiently acute for correlations to occur (if not—if the angle is too large—then, as is well known empirically, each side corresponds to single-particle interference, and the correlations disappear).[32] Also, in the space of all paths a tangible particle always takes a single trajectory by RPP in Sec. IIA. This means that there are just two choices for what we term the "upper" tangible particle: The upper tangible can only take an upper path—that is, either a path in the homotopy class $A$ or in the class $B'$ in the three-dimensional setup on the right in Fig. 3—which path it does take varies randomly by RPP at each trial of the experiment. If the upper tangible takes, say, a path in *A* then the other, lower tangible must by conservation of momentum take a linked, oppositely directed path in *A´*. Similarly if the upper takes a path in *B´* then the other tangible is linked to take a path in *B*.

For simplification, and assuming a fixed time interval, we will now introduce standard Dirac notation for the amplitudes involved (although of course the underlying space remains nonmeasurable). That is, we let

$$\langle C|A\rangle = \sum_{\substack{\text{over all upper paths} \\ x(t) \text{ in } A \text{ to } C}} \exp(iS[x(t)]/\hbar), \text{ and } \langle C|B\rangle = \sum_{\substack{\text{over all lower paths} \\ x(t) \text{ in } B \text{ to } C}} \exp(iS[x(t)]/\hbar).$$

Similarly for $\langle C´|B´\rangle$ and $\langle C´|A´\rangle$.

We note further that in SP when a tangible particle goes from a beamsplitter to a detector in our interferometer its destination is already determined by interference effects at the beamsplitter (a postulate, if you will), which in turn is a consequence of the exponentiated actions associated with the particles traveling the paths from the source to the beamsplitter (the sum of unit vectors as in Fig. 1). Moreover, the particles taking paths in the homotopy classes *A* and *A´* are from different streams; likewise for those taking paths in *B* and *B´*. By this last fact (according to NIP in Sec. II.B), we obtain the combined amplitude by multiplying the separate ones. Since we have two cases, we then sum the results:

$$\langle C|A\rangle\langle C´|A´\rangle + \langle C|B\rangle\langle C´|B´\rangle. \tag{2}$$

Now the amplitude in expression (2) is just the usual SQM expression for particles in the singlet state. In Schrödinger's definition, it comes from an "entangled state," one that cannot be factored into states coming separately from the tensor product spaces. Indeed, as it stands, expression (2) seems to suggest that the particles must somehow communicate across the source in order to coordinate the predicted correlations.

But wait! Here is where shadow particles and congruent paths make all the difference. Observe that each lower path on the left side is congruent to a lower path on the right side, and so $\langle C|B\rangle = \langle C´|A´\rangle$. Thus substituting equals for equals in expression (2) we have,

$$\langle C|A\rangle\langle C´|A´\rangle + \langle C|B\rangle\langle C´|B´\rangle = \langle C|A\rangle\langle C|B\rangle + \langle C´|B´\rangle\langle C´|A´\rangle. \tag{3}$$

Now assume, say, $\alpha \geq \beta$ (the other case is similar). Note that each upper path on the left side (that is, in *A* from S to *C*) has a sub-path congruent to a corresponding upper path on the right side (a path in *B´* from S to *C´*); and the action over the remaining small portion of the path in *A* to *C* has an exponentiated-action proportional to $e^{i(\alpha-\beta)}$. That is, the sum of exponentiated-action terms over all upper paths traveled by the stream of particles on the left side, $\langle C|A\rangle$, is proportional to $e^{i(\alpha-\beta)}\langle C´|B´\rangle$, where $\langle C´|B´\rangle$ is the sum of exponentiated-action terms over all upper paths on the right side. From Eq. (3) and $\langle C|B\rangle = \langle C´|A´\rangle$ we therefore have,

$$\underbrace{\langle C|A\rangle\langle C|B\rangle}_{\text{left side}} + \underbrace{\langle C'|B'\rangle\langle C'|A'\rangle}_{\text{right side}} \propto e^{i(\alpha-\beta)}\langle C'|B'\rangle\langle C'|A'\rangle + \langle C'|B'\rangle\langle C'|A'\rangle. \tag{4}$$

$$= (e^{i(\alpha-\beta)} + 1)\langle C'|B'\rangle\langle C'|A'\rangle \propto 2(e^{i(\alpha-\beta)} + 1)e^{i\beta} = 2(e^{i\alpha} + e^{i\beta}) \tag{5}$$

When we multiply the right-most expression in Eq. (5) by its complex conjugate, we obtain a result proportional to the standard probability that both detectors will fire:

$$\cos^2\frac{\alpha-\beta}{2}. \tag{6}$$

This is the usual result. But, as noted in Sec. II.E, the calculation in SQM is carried out in an infinitely-many-to-one image of SP. It predicts the observable with great accuracy, but offers no insight into the underlying causative factors. This is not the case in SP. In fact Eq. (4) is reminiscent of what Bell (above) termed "locally explicable," involving an "uncoupling of the fluctuations" as in $P(\alpha,\beta|\lambda) = P(\alpha|\lambda)P(\beta|\lambda)$.

A local theory is necessarily deterministic.[33] Bell often likened this determinism to the behavior of identical human twins carrying with them identical genetic instructions.[34] In our interferometer the instructions take the following form: The sum over the collective, exponentiated-actions (the "action instructions") in effect instructs the particles where to go when phase shifts are performed on each side. The particles don't somehow have to know what occurs on the other side. Rather, as inanimate entities, they blindly pursue all possible paths on their way from the source to a beamsplitter. The outcome (which detectors fire) is simply the result of the particles' built-in programming to follow the rules of physics pertaining to the addition of exponentiated-action terms, where the action rules are equivalent to Newton's laws.[35] As in Fig. 1, the constructive and destructive interference effects result from sums of infinitely many unit vectors in the complex plane—that is, from sums over exponentiated-action terms corresponding to paths traveled within the interferometer from the source to a beamsplitter. When particles (tangible or shadow) travel congruent paths they have identical actions, and therefore have identical exponentiated-action terms that contribute to the sum in the propagator.

When the congruent paths common to each side of the setup are taken into account, Eq. (4) exhibits, as in Bell's Lille-Lyons example, independent, decoupled interaction on each side. The common causative factors, indicated in Eq. (4), correspond to the congruent paths common to the two sides. Eq. (4) predicts the correlations: correlations occurring 100 percent of the time when the phase shifts $\alpha$ and $\beta$ are the same, and occurring with a frequency according to expression (6) otherwise.

In SP the underlying mechanism comes from shadow particles traveling paths not taken by the tangibles. It has nothing to do with information transmitted across the source. Rather it has to do with the particles possessing identical action instructions at the source. In discussing single-particle interference the mathematical physicist Roger Penrose once wrote:[36]

> "… the key puzzle is that somehow a photon (or other quantum particle) seems to have to 'know' what kind of experiment is going to be performed upon it well in advance of the actual performing of that experiment. How can it have the foresight to know whether to put itself into 'particle mode' or 'wave mode' as it leaves the (first) beamsplitter."

But particles are inanimate objects that know nothing. How can the particles in a two-particle experiment know where to go? They don't know, of course. As noted above, they mindlessly follow their action instructions. As always in SP (whether for single- or multi-particle interference), the particles in a stream travel all possible paths blindly from the source to a beamsplitter, but do so obeying their built-in instructions. In a two-particle experiment the correspondingly congruent paths between the two sides lead to the predicted correlations. When the upper and lower shadow streams converge at a beamsplitter, there is a local, contiguous sharing of information. The local interference effects (the sum of unit vectors, as in Fig. 1) determine where the tangible particles go next. There is no collapse of a wavefunction, as is necessary in SQM. Nothing needs to be sent across the origin in SP, contrary to what appears to be the case if the hidden reality of shadow particles pursuing all possible paths is not taken into account. The result is an entirely locally explicable process—via local, contiguous interference effects. There is no action at a distance, although the end result is as though there were. **QED**

In a moment we shall carry out two similar arguments, one for entangled quantum beads on a ring, and one for spin. But in a strict logical sense these are unnecessary, since it only takes one counterexample to overturn a theorem. Still, for historical reasons, the argument for spin has some interest in its own right, because in the original 1964 proof of his theorem Bell had in mind David Bohm's spin version of a gedanken experimental test of the EPR correlations.[37] The argument for entangled beads (the next counterexample) prepares the way for the spin argument.

**B. Counterexample 2: Entangled ring particles**

As we saw in Sec. III.B.2 the associated configuration space for the particle-on-a-ring is not simply connected and the overall path-propagator is a linear combination of propagators over each homotopy class in the space. Thus the propagator for the ring over a fixed time interval and for $0 \leq \alpha, \theta \leq 2\pi$ is

$$K(\alpha,\theta) = \sum_{n=-\infty}^{\infty} K_n(\alpha,\theta) \tag{8}$$

where $K_n(\alpha,\theta)$ is proportional to the sum of all exponentiated-action terms for paths around the circle in the homotopy class corresponding to the winding number, $n$ (the action for the propagator is just that of a free particle).[38]

Now—in an instructive but doubtless experimentally infeasible gedanken experiment—consider a pair of entangled particles emitted by a source event and traveling in opposite directions along a pair of rings that are somehow separated and moved apart following the source event. We will assume that the particles receive the same initial momentum kick, starting from some initial point $\theta$ on the associated initial pair of circles. The two tangible particles will then, randomly, travel in a pair of corresponding homotopy classes (by conservation of momentum), which we denote by $n$ and $-n$. This is a local event occurring at the source, a consequence of the initial input of momentum. Assume now that a momentum measurement at angle-point $\alpha$ is made on the left circle and one at $\beta$ is made on the right. By RPP and NIP above, we see that the combined amplitude for this is

$$\sum_{n=-\infty}^{\infty} K_n(\alpha,\theta) K_{-n}(\beta,\theta) \tag{9}$$

The propagators in the sum in expression (9) contain the information that gives the time-evolution of the quantum system for the entangled ring particles.

In one of several cases involving the relative positions of the angle-points on the two circles, assume that $0 \leq \theta \leq \beta \leq \alpha < \pi$ (the arguments for the other cases are similar). Now it is already intuitively clear that the propagator $K_n(\alpha,\theta)$ dealing with the left side of the experimental arrangement contains the information of the propagator $K_{-n}(\beta,\theta)$ on the right side, since the set of all paths from $\theta$ to $\alpha$ includes those from $\theta$ to $\beta$. Thus the product of amplitudes in expression (9) is "almost completely" based on paths that on each side of the experimental setup are congruent to corresponding ones on the other side. But a simple mathematical argument below brings out the correspondence more precisely as in expression (7) above.

Consider any path from $\beta$ to $\alpha$ of winding number 0. This path, in traveling from $\beta$ to $\alpha$, can loop any number of times in a given direction around the reference point 0 on the circle, but it must always in its overall passage around the circle travel (but not in any particular sequence) the same number of times past 0 in the opposite direction. Hence, for each $n$, any path from $\theta$ to $\alpha$ of winding number $n$ is homotopically equivalent to one from $\theta$ to $\beta$ of winding number $n$ plus one from $\beta$ to $\alpha$ of winding number 0.

As is well known, the exponentiated action for a free particle traveling around the circle on any path from $\beta$ to $\alpha$ of winding number 0 is a function $f(\alpha,\beta)$ whose value given a particular input of momentum is proportional to $e^{i(\alpha-\beta)}$.[39] One can see this intuitively, since the exponentiated action for paths corresponding to congruent but oppositely directed paths around the circle from $\beta$ to $\alpha$ will cancel, leaving just the action over the minimal path from $\beta$ to $\alpha$. Thus we have

$$\sum_{n=-\infty}^{\infty} K_n(\alpha,\theta)K_{-n}(\beta,\theta) = f(\alpha,\beta) \sum_{n=-\infty}^{\infty} K_n(\beta,\theta)K_{-n}(\beta,\theta). \tag{10}$$

Thus in the case we are considering ($0 \leq \theta \leq \beta \leq \alpha < \pi$), the amplitude for finding correlated momentum observables involves a function that depends locally on the least path from $\beta$ to $\alpha$.
**QED**

### C. Counterexample 3: Entangled spin1/2 particles

We shall reduce the argument for this counterexample to that of the previous one for ring particles in the following way:

First, note that the paths for spin in SP are "rotational" paths in $\mathbb{R}^3 \times SO(3)$. As is well known, we can model $SO(3)$ homeomorphically in terms of a solid ball of radius $\pi$, with antipodal points identified. In this picture, a point a distance $\gamma$ from the center of the ball and along the radius pointing in the direction of the unit vector $\mathbf{n}$ corresponds to a counterclockwise rotation $\gamma$ about the axis $\mathbf{n}$. Because $SO(3)$ has only two homotopy classes of paths and because we are identifying antipodal points on the ball, it is an interesting fact that a path along, say, the $y$-axis that passes just once through the outer surface of the ball is not homotopic to a path that does not pass through at all, although paths that pass through any even number of times are so homotopic. To see this, suppose a path along the $y$-axis runs from $(0, 0, 0)$ (the center of the ball) in the positive direction to the point $(0, \pi, 0)$ on the ball) and then jumps back to the antipodal point $(0, -\pi, 0)$, and continues on to do this a second time, so that it runs from 0 to $4\pi$ along the y-axis. This generates a closed loop in which the paths can be moved continuously to the ball's surface, still connecting $(0, \pi, 0)$ to $(0, -\pi, 0)$ twice. Now mirror the second half of the path to the antipodal side, which produces a closed loop on the surface of the ball connecting the point $(0, \pi, 0)$ on the y-axis to itself along a great circle. This circle can be shrunk to a point. We will take this into account below, when we define a kind of winding number for paths looping back and forth along the y-axis.

Now consider a pair of entangled spin-1/2 (tangible) particles traveling in opposite directions along, say, the y-axis to SG devices on each side of an experimental setup. As we have emphasized, in SP a path is a continuous sequence of rotations in the space $SO(3)$ (that is, in $\mathbb{R}^3 \times SO(3)$). Now assume, for convenience, that both tangibles after they leave the source event are passed through a filter oriented at angle $\theta$ along the y-axis. Assume also that a measurement of angle $\alpha$ is made on the left side about the y-axis, while one of $\beta$ is made on the right, where (as in B, counterexample 2) $0 \leq \theta \leq \beta \leq \alpha < \pi$. Thus, as noted in the discussion of expression (8) above, the propagator corresponding to paths from $\theta$ to $\alpha$ on the left side of the experimental setup contains the information of all the paths going from $\theta$ to $\beta$ on the right, since the set of all paths from $\theta$ to $\alpha$ includes those from $\theta$ to $\beta$. Moreover, we shall conclude as follows (as in B, counterexample 2) that there is a formula expressing this information.

In computing the propagator from $\theta$ to, say, $\gamma$ along the y-axis in the ball of radius $\pi$ (that is, a rotation of angle $\gamma$ about the y-axis) the path integral sums the exponentiated-action terms over *all* possible paths from $\theta$ to $\gamma$ in the ball. But, in a key simplification for our discussion, we can assume dominance of rotations about the y-axis (so what we have here is an approximation). Thus assume that the SG field is titled at some angle about the y-axis and that the tangible on the left side is traveling in the $SO(3)$ homotopy class $A$ and the other tangible in $A´$ on the right.

Hence, on each side we picture the tangible's path as being projected onto a continuous path along the y-axis of the ball. In going from $\theta$ to another angle-point $\gamma$, such a path *confined* to the y-axis, can travel any number of times back and forth, sometimes passing through the ball. This will enable us to carry out a construction similar to that of the ring counterexample above, but where we can ignore the difference in the actions involved in the two contexts. (As mentioned, the action for spin is the sum of a spinning top component and a magnetic field component, but this detail will not affect our argument).[40] Also, to avoid the fact noted above about paths passing through the ball an even number of times etc., we constrain the paths to the y-axis. Thus a path is "stuck" to the y-axis and can no longer be moved continuously from the y-axis up to the surface of the ball, which eliminates any homotopy involving an even number of passages through the surface. This enables us to speak of winding numbers $n$ about the reference point 0 for a path that goes from $\theta$ to $\gamma$ and possibly passes through the ball as the path runs back and forth along the y-axis.

By this construction of projecting onto the y-axis in the ball (corresponding to our SG measurements about the y-axis on each side of the experimental arrangement) we thus generate a set of homotopy classes different from the conventional two given by $SO(3)$ (the usual two can be pictured by two regions on a two-sphere [41]). This enables us to reduce everything to the previous counterexample. Therefore when we sum over these homotopy classes labeled $n$ we obtain a result similar to EQ (8), where the difference between the sides is just a function of $\alpha$ and $\beta$. For spin this means that the common information between the sides comes from correspondingly congruent stream paths. Thus, as with the previous two counterexamples, congruent paths on each side of the experiment lead to the correlations when SG measurements are performed on the spatially separated sides. The correlations occur by means of initial information imparted at the source to the shadow streams—that is, the correlations are predetermined at the source as a function of the subsequent measurements. The interactions producing the correlations are purely local throughout the experiment. Nothing needs to be sent across the origin. **QED**

**VI. CONCLUSION**

Einstein stated in a 1948 letter to Max Born: "… I still cannot find any *fact* [italics added] anywhere which would make it appear likely that [this] requirement [the independent existence of physical reality in different parts of space] will have to be abandoned. I am therefore inclined to believe that the description of quantum mechanics … has to be regarded as an incomplete and indirect description of reality …." [42] It has been long and widely believed that the "fact" unknown to Einstein was provided by Bell's theorem—which, as we have now shown, fails in the space of all paths.

Still, one can always reject shadow particles by arguing that they can be known only indirectly. But, interestingly, it turns out that either we must choose shadow particles or, alternatively, "spooky action at a distance." These are the two mutually exclusive choices for quantum reality (in a way perhaps the most important result in this paper). This conclusion is based (somewhat ironically given the original intent of the authors) on an obvious and straightforward extension of an argument by Bernstein, Green, Horne, and Zeilinger (BGHZ). [43] As mentioned in Sec. V, BGHZ had to augment their premises by what they called "emptiness of paths not taken," EPNT. They then proved that if the possible paths not traveled by the entangled tangible particles are actually empty, then the quantum correlations must occur through "action at a distance."

BGHZ based their argument on a two-particle interferometer, but their proof can be extended to any multiply-connected quantum system (the two-particle interferometer in Sec. V.A is of course an elementary example of such a system; and spin spaces in SP are also multiply connected). Indeed, by assuming that the paths not traveled by the tangible particles are empty then the straightforward BGHZ argument applies, *mutatis mutandis*, to homotopy classes of paths in a general space in place of the small number of paths analyzed by the BGHZ paper. In other words, we have the result that "SP" implies "no action at a distance," and "not SP" implies "action at a distance."

Now one could deny EPNT and still have a nonlocal theory (for example, David Bohm's pilot-wave theory, which in effect "fills" the paths with a "pilot wave").[44] Also there are partial forms of SP that claim to be local or realistic theories—for example, "consistent histories"(CH).[45] CH is a mathematically sound theory based on SQM (modifying an approach to quantum logic developed by Birkhoff and von Neumann). [46] But, because it is ultimately based on SQM, it follows that in CH such terms and concepts as "and," "or," and "local" no longer have their ordinary, commonsense meanings unless they are used according to CH's "single framework rule." This rule allows one to employ standard Boolean logic and concepts such as "local" only on single frameworks or histories (path sequences that do not involve non-commuting quantum operators along the sequence). This rules out as meaningless (in the sense of "speak no evil") any discussion of what appear, from a commonsense standpoint, to be the nonlocal experimental outcomes implied by Bell's theorem. Yet the single framework rule is in fact merely a grammatical or syntactical rule, although the grammar is based on the properties of operators in the Hilbert space of SQM— which as we have noted is just an infinitely-many-to-one image of SP. Indeed, this syntactical rule leads to a picture of reality in which infinitely many separate slices or histories can be true, although for infinitely many of the pairs of histories taken together, any comparison between them would be

18
meaningless.[47] CH maintains that this paradox is not a defect in the theory, but merely reflects the way the quantum world happens to be.

Perhaps, but there are no such paradoxes associated with SP. In SP "locality" has its usual meaning of contiguous interaction, and everything is an elegant sum of classical dynamics over the paths in a shadow stream. In EPR's terms above, it is a compete theory with each element of reality having a counterpart in the theory. We can freely combine logical statements within the theory using classical Boolean logic. Thus, like special relativity, for example, the space of all paths is straightforward and easy to grasp and conforms perfectly with informed commonsense (that is everything follows logically from the fundamental properties).

Note also, contrary to what is commonly supposed in these matters: One cannot reject the space of all paths on the basis of Occam's razor ("entities should not be multiplied beyond necessity"). The reason is that Occam's razor (as stated) is vague and imprecise about the criteria for its strictures. It can, however, be expressed precisely in complexity theory and given a form amenable to mathematical analysis: "If presented with a choice between indifferent alternatives, then one ought to select the simplest one."[48] Put this way, the alternatives here are decidedly not indifferent: The non-existence of shadow particles leads to paradox and, in the case of entangled particles, to something close to an absurdity that not only conflicts with special relativity but would also have to take place according to utterly incomprehensible dynamical laws.

One conceivable difficulty with SP is, "What happens to the shadow energy as particles travel their respective paths?" But, as generally happens with Feynman's formulation, the sum of interference effects handle this via cancellation of associated unit vectors except in the neighborhood of the classical path, as in Fig. 1. The SP universe is, in fact, essentially equivalent to that of MWI, different worlds corresponding to different paths, the paths generated in a quantum evernt never intersecting, although frequently infinitely close.

Now shadow particles and their bizarre implication of infinitely many parallel patterns of matter (that is, parallel copies) interacting with our tangible universe may be novel and at first seem contrary to our everyday experience (because they constitute a hidden reality), but they are not absurd: Outcomes in SP are simply a sum over classical paths, where all particles in a stream obey classical dynamics. Of course shadow particles can be known only indirectly at present; but, as we know, this has happened before in science. For example, as far back as 60 BC the Roman poet Lucretius (in his amazingly prescient poem *On the Nature of Things*) hypothesized the existence of atoms on the basis of what we now call Brownian motion—in the form of the bombardment of "dancing" dust particles by "hidden" atoms. And later, in the nineteenth century long before more direct experimental confirmation was feasible, many physicists (again thinking in terms of the indirect evidence generated by Brownian motion) conjectured the existence of atoms and molecules. Something similar is occurring with Deutsch's shadow particles. To reject them is to embrace action at a distance—which in the words of Einstein *et al*. "no reasonable definition of reality could be expected to permit …").[49]

As Bell once stated to Bernstein, "The discomfort that I feel [regarding the implications of his theorem] is associated with the fact that the observed perfect quantum correlations seem to demand something like the "genetic" hypothesis [identical twins, carrying with them identical genes]. For me, it is so reasonable to assume that the photons in those experiments carry with them programs, which have been correlated in advance, telling them how to behave. This is so rational that I think that when Einstein saw that, and the others refused to see it, *he* was the rational man. The other people, although history has justified them [contrary to the conclusion of the present paper], were burying their heads in the sand. I feel that Einstein's intellectual superiority over Bohr, in this instance, was enormous; a vast gulf between the man who saw clearly what was needed, and the obscurantist. So for me, it is a pity that Einstein's idea doesn't work. The reasonable thing just doesn't work."

Ah, but it does.